# Superradiance and subradiance in extended media

Jonathan F. Schonfeld

Harvard-Smithsonian Center for Astrophysics, Cambridge, Massachusetts 02140, USA

In super- or subradiance, a quantum superposition of excited atoms collectively emits a photon much more or much less rapidly than an isolated atom. Superradiant and subradiant lifetimes have been derived for finite spheres of uniform media, either by simulating random samples or by expanding in spherical harmonics and analyzing Bessel functions. We introduce a simple regulator that applies to unbounded media, enabling trivial derivation and analysis of lifetimes via elementary Fourier techniques. The regulator can be interpreted as a correlation length for atomic positions; the regularized system describes localized regions of enhanced radiative activity in otherwise quiescent surroundings.

## I. INTRODUCTION

Superradiance and subradiance [1] play prominent roles in quantum optics. Quantitative characteristics of superradiant and subradiant decay have been worked out in considerable detail [2-6] for photons idealized as scalar quanta. For coherent superpositions of single excited atoms in fixed positions with number density $\rho$, the decay rates $\lambda$ (normalized to isolated-atom decay rate $\gamma$) and amplitude patterns $\beta$ are given [2] by eigenvalues and eigenvectors of

$$\int d^3\mathbf{r}' \, \rho(\mathbf{r}')K(\mathbf{r}-\mathbf{r}')\beta(\mathbf{r}') = \lambda\beta(\mathbf{r}), \qquad (1)$$

where

$$K(\mathbf{r}-\mathbf{r}') = \frac{\exp(ik_0|\mathbf{r}-\mathbf{r}'|)}{ik_0|\mathbf{r}-\mathbf{r}'|} \qquad (2)$$

and $k_0$ is the wavenumber of resonant photon emission. This system has been studied in considerable detail for uniform finite spheres. Lifetimes have been obtained by simulating random samples [5] or by expanding eigenvectors of Eq. (1) in spherical harmonics and solving complicated auxiliary equations that involve Bessel functions [2,3]. Accurate eigenvalues require numerical evaluation, but in general, for a sample of large radius $R \gg k_0^{-1}$, all but $k_0^2 R^2/\pi$ modes have very small (subradiant) real parts and the eigenvalues of the remaining modes all have roughly the same (superradiant) real part

$$\frac{N}{k_0^2 R^2/\pi} = \frac{4\pi^2}{3}\frac{R\rho}{k_0^2}, \qquad (3)$$

where $N$ is the total number of atoms in the sample. Near the superradiant modes, the real and imaginary parts of the lifetime exhibit resonant behavior as a function of a suitable mode index.

For very small $R$ (Dicke limit [1]), all modes but one are subradiant, and the superradiant eigenvalue has real part $N$.

In this paper we show how to replace the finite radius $R$ with a simple correlation-length regulator that applies to unbounded media ($R\to\infty$). This makes derivation of lifetimes mathematically trivial by enabling use of elementary Fourier techniques. It also addresses an unphysical aspect of Equation (3): Equation (3) says superradiant decay rates diverge (and the corresponding eigenstates extend over the entire medium) as sample size goes to infinity. This entails an unreasonable degree of cooperation across large distances. The correlation length resolves this difficulty. We will not attempt to estimate correlation lengths (if applicable) in real media.

The remainder of this paper is organized as follows. In Section II we introduce the correlation length and derive the modified decay spectrum for unbounded media. In Section III we check the result by comparing it with the mode-counting rules of thumb in the discussion surrounding Equation (3). We summarize in Section IV.

## II. UNBOUNDED MEDIUM

The right-hand side of Equation (2) is simply a Yukawa potential with imaginary mass, so for $R\to\infty$ the formal solution to Equation (1) for wavevector $\mathbf{k}$ is

$$\lambda = \frac{4\pi\rho/ik_0}{k^2 - k_0^2}. \qquad (4)$$

As expected from finite $R$, this is purely imaginary (no decay, i.e. subradiance) except at the singularity, where it has infinite real principle part.

In practice, a finite correlation length regularizes the singularity. To motivate the specific form of regularization, consider that the kernel $K$ originates [3] in the product of photon field strength and its conjugate at two atomic locations,

$$\exp(i\mathbf{k}\cdot(\mathbf{r}_a - \mathbf{r}_b)) \equiv \exp(i\mathbf{k}\cdot\Delta\mathbf{r}). \qquad (5)$$

$K$ is a weighted integral of this expression over the components of wavevector $\mathbf{k}$. At least for a solid atomic medium, where nearby atoms can retain their relative positions for extended times. For larger and larger $\Delta\mathbf{r}$ it is less and less likely that points a and b maintain the same phase relationship for the entire duration of the emission process. To quantify this effect it seems reasonable to replace Expression (5) with

$$\exp(i\mathbf{k}\cdot\Delta\mathbf{r})\exp(-\mu|\Delta\mathbf{r}|) \qquad (6)$$

for some phenomenological parameter $\mu$, where $\mu^{-1}$ is the correlation length. If for practical purposes $\mu$ is insensitive to $\mathbf{k}$, then Equation (2) will become

$$K(\mathbf{r}-\mathbf{r}') = \frac{\exp((ik_0-\mu)|\mathbf{r}-\mathbf{r}'|)}{ik_0|\mathbf{r}-\mathbf{r}'|} \qquad (7)$$

and Equation (4) becomes

$$\lambda = \frac{4\pi\rho/ik_0}{k^2 + (\mu - ik_0)^2}. \qquad (8)$$

For $\mu < k_0$, the real part of this function has peak value

$$\lambda_{\text{peak}} = \frac{2\pi\rho}{\mu k_0^2} \qquad (9)$$

on the sphere of radius

$$k = (k_0^2 - \mu^2)^{1/2} \qquad (10)$$

and resonant behavior about this peak is obvious using $k$ as mode index. The thickness of the peak about the surface of this sphere is $\sim 2\mu$ for small $\mu$. Thus, in this case a superradiant state is an excitation localized to a region of linear size $O(\mu^{-1})$ in position space, having atomic amplitudes modulated with spatial frequency $k_0$.

For $\mu > k_0$, the real part of Eq. (8) has peak value

$$\lambda_{\text{peak}} = \frac{8\pi\rho\mu}{(\mu^2 + k_0^2)^2} \approx \frac{8\pi\rho}{\mu^3} \qquad (11)$$

at $k = 0$, where the approximate equality holds for large $\mu$ (Dicke limit), and the width of the peak is roughly $\mu$. In this case a superradiant state is an excitation localized to a region of linear size $O(\mu^{-1})$ in position space, with negligibly modulated atomic amplitudes.

### III. MODE COUNTING

As a sanity check, we compare the results of the preceding section with the mode-counting statements from Section I.

If the (fictitious) quantization volume is a cube of side $L$, then the total number of modes in the superradiant shell in **k**-space is, for small $\mu$,

$$\frac{(2\mu)(4\pi k_0^2)}{(2\pi/L)^3} = \frac{\mu k_0^2 L^3}{\pi^2}. \qquad (12)$$

Therefore the total number of superradiant modes in a superradiant volume of side $\mu^{-1}$ in position space is

$$\frac{k_0^2}{\pi^2 \mu^2}. \qquad (13)$$

According to Equation (3), if we divide this into the total number $\rho\mu^{-3}$ of atoms in the superradiant volume, we should get the peak decay rate, Equation (9), possibly up to a numerical factor reflecting differences between spherical and cubic geometries. Indeed,

$$\frac{\rho\mu^{-3}}{k_0^2/\pi^2\mu^2} = \frac{\pi}{2}\lambda_{\text{peak}}. \qquad (14)$$

For large $\mu$ (Dicke limit), the superradiant region in **k**-space is a sphere of radius $\mu$, and therefore volume $4\pi\mu^3/3$. Therefore the total number of superradiant modes in a spatial cube of side $\mu^{-1}$ is $1/6\pi^2$ independent of $\mu$. According to the discussion in Section I, if we divide this number into the total number $\rho\mu^{-3}$ of atoms in the superradiant volume, we should get the peak decay rate, Equation (11), possibly up to a numerical factor reflecting differences between spherical and cubic geometries. Indeed,

$$\frac{\rho\mu^{-3}}{1/6\pi^2} = \frac{3\pi}{4}\lambda_{\text{peak}}. \qquad (15)$$

### IV. SUMMARY

We have generalized the theory of superradiant and subradiant decay to unbounded media by introducing a regularizing parameter with a straightforward physical interpretation. This makes the formal mathematics trivial and enables consideration of localized superradiance in otherwise quiescent surroundings.